\documentstyle[12pt]{article}
\begin{document}
\begin{titlepage}
\begin{flushright}
CERN-TH.6434/92\\
ENSLAPP-A-377/92
\end{flushright}
\vfill
\begin{center}
{\Large \bf Comment on
``High Temperature Fermion Propagator --
Resummation and Gauge Dependence of the Damping Rate''}\\
\vfill
{\large Anton Rebhan}\\
\bigskip
{\sl
Laboratoire de Physique Th\'eorique ENSLAPP\footnote{URA 14-36
du CNRS, associ\'ee \`a l'E.N.S. de Lyon, et au L.A.P.P.
d'Annecy-le-Vieux}\\
BP 110, F-74941 Annecy-le-Vieux CEDEX, France\\}
\medskip
and\\
\medskip
{\sl
Theory Division, CERN\\
CH-1211 Geneva 23, Switzerland\\}
\vfill
{\large ABSTRACT}
\end{center}
\begin{quotation}
Baier et al.~have reported the damping rate of long-wavelength
fermionic excitations in high-temperature QED and QCD to be
gauge-fixing-dependent even within the resummation scheme
due to Braaten and Pisarski.
It is shown that this problem is caused by the singular
nature of the on-shell expansion of the fermion self-energy
in the infra-red. Its regularization reveals that the alleged
gauge dependence pertains to the residue rather than the pole of
the fermion propagator, so that in particular
the damping constant comes out gauge-independent, as it should.
\end{quotation}
\vfill
\begin{flushleft}
ENSLAPP-A-377/92\\
CERN-TH.6434/92\\
March 1992
\end{flushleft}
\end{titlepage}

During the last decade, there has been much controversy about
perturbative QCD at high temperature, nurtured by contradictory
results on the leading-order damping rates of plasma excitations,
which, at the one-loop order, came out gauge-dependent in both
magnitude and sign [\ref{rPr}]. It had been surmised rather early
that higher-loop orders are relevant for a complete calculation
of the leading-order damping rates, but only rather recently has
it been established that indeed a proper resummation cures both
the reported gauge dependences and the wrong sign found in
most one-loop calculations. On the one hand, Kobes, Kunstatter, and
the present author [\ref{rKKR}]
have proposed a formal argument [\ref{fKKR}] verifying the
gauge independence of plasma dispersion relations once
they have been evaluated
completely, and, on the other hand, Braaten and Pisarski
have developed a systematic procedure for a resummation of
leading-temperature corrections (``hard thermal loops'') into an
effective loop expansion where higher-loop orders are also of higher
order in the coupling constant [\ref{rBP}].

The whole issue is in fact not peculiar to nonabelian gauge theories,
but can be studied already in QED, where the calculation of the
damping rate of fermionic excitations equally requires resummation
of all hard thermal loops.

The potential gauge dependence of the leading-order damping rate
of long-wave\-length fermionic excitations in high-temperature QED
(or QCD) within the resummation scheme of Braaten and Pisarski has
been scrutinized most recently by Baier, Kunstatter, and Schiff
[\ref{rBKS}]
with the surprising conclusion that there were in fact
gauge-dependent contributions to the imaginary part
of the high-temperature fermion propagator at order $g^2T$.

With $\Delta_f$ being the fermion propagator
that resums the contributions
of order $gT$, the part of the fermion self-energy
proportional to the gauge parameter $\xi$,
at order $g^2T$, reads [\ref{rBKS}]
$$\delta\Sigma(P)=\xi g^2 \Delta_f^{-1}(P)
\int_\beta{d^4K\over(2\pi)^4}{1\over K^4}
\left[ \Delta_f(P-K) \Delta_f(P)^{-1} -1 \right]. \eqno(1)$$
While gauge dependence appears to hold formally
``on-shell'' ($\Delta_f^{-1}
=0$) in accordance with the analysis of Ref.~[\ref{rKKR}],
the thermal integral in (1) develops on-shell poles which
just cancel the explicit factors $\Delta_f^{-1}$,
giving a non-zero, gauge-dependent
imaginary contribution to the on-shell
value of the fermion self-energy, hence to the damping rate!

In order to highlight
what exactly goes wrong
in this calculation, it is sufficient to consider
a partial resummation in the
long-wavelength limit, where only
a constant thermal mass is taken
into account [\ref{rKKRS}] by a
resummed lowest-order Lagrangian for the fermions
$${\cal L}=
i\bar\psi {D}\llap{/} {D^2+m^2\over D^2} \psi, \eqno(2)$$
where $m\sim gT$ is the thermal mass of fermionic modes in the
long-wavelength limit. Since there appeared to be no problem with
the contribution from the cuts of the full propagator, which are
the main omission in (2), the same problem will show up in this
simpler setting, together with its solution.

The gauge-invariant form of (2)
guarantees that the resummed terms obey simple Ward identities,
so that the gauge dependence of the fermion self-energy is
again given by (1), which now more explicitly reads
$$\delta\Sigma(P)\bigg|_{{\bf p}\to0}=\xi g^2\gamma_0
\left[ {\delta^2\over p_0^2}
\int_\beta{d^4K\over(2\pi)^4}{1\over [K^2]^2}
{p_0-k_0\over(P-K)^2-m^2}-{\delta\over p_0}
\int_\beta{d^4K\over(2\pi)^4}{1\over [K^2]^2} \right], \eqno(3)$$
where $$\delta\equiv P^2-m^2.$$
(Here the notation for 4-vectors is $K=(k_0,{\bf k})$, with
metric $(+,-,-,-)$;
$\int_\beta$ refers to evaluation at finite temperature.)

The usual trick to evaluate the double pole
$1/(K^2)^2$ in (1) is to write
$${1\over [K^2]^2}={\partial\over\partial\lambda}
{1\over K^2-\lambda}\bigg|_{\lambda=0}. \eqno(4)$$
It is easy to see that, after
this replacement, the integrals in (3)
could not have any imaginary
part on the ``mass-shell'' [$\delta=
(p_0^2-m^2)=0$] for all $\lambda>0$,
owing to the lack of phase-space
volume for the corresponding decay process. By continuity and
intuition, one therefore would not expect any contribution to
the damping rate from expression (3), the more so because
the integrals
in (3) are multiplied by explicit on-shell zeros. 
Yet, what the authors of Ref.~[\ref{rBKS}] have found is that
evaluating the integrals off-shell and setting $\lambda=0$
after differentiation according to (4)
does give rise to an imaginary
part, which has a double pole on-shell, exactly compensating the
on-shell zeros in $\delta\Sigma$. This leaves a non-vanishing,
gauge-dependent contribution to the imaginary part
of the dispersion relation!

Indeed, the order
$$\lim = \lim_{p_0\to m} \lim_{\lambda\to0} \eqno(5)$$
chosen in Ref.~[\ref{rBKS}] seems perfectly
natural, if the introduction of $\lambda$ was just
a technical device for the evaluation
of the double pole $1/(K^2)^2$
[\ref{fG}] [\ref{rNNP}].

Still, the discontinuous behaviour with respect to this parameter
$\lambda$, which looks suspiciously like an infra-red regulator,
calls for a closer look at the infra-red behavior of the integrals
in (3). Taking the order of the limits (5) for granted, the latter
can be studied in a most clear-cut way by the introduction of an
actual infra-red cut-off of spatial momenta, $|{\bf k}| \ge \mu$.
Straightforward evaluation of
the leading-temperature contributions to (3) yields
$$
\delta\Sigma(P)\bigg|_{{\bf p}\to0}=
\xi g^2 \gamma_0 {T\over2\pi^2} {\delta\over p_0}
\left( 1-{3\delta\over4p_0^2}+{\delta^2\over4p_0^4}+
  {\delta\over2}\Bigl(1-{\delta\over2p_0^2}\Bigr)^2
  {\partial\over\partial\delta} \right) I,$$
$$ I=
\int_\mu^\infty {dk\over k^2-\delta^2/4p_0^2}
={p_0\over\delta}\ln{\mu+\delta/2p_0\over\mu-\delta/2p_0}.
\eqno(6)$$

Without an infra-red cut-off ($\mu\to0$), the logarithm in (6)
becomes $i\pi$, and apparently a finite gauge-dependent
imaginary contribution is added to the on-shell pole,
which, basically, is the result found in Ref.~[\ref{rBKS}].
However, keeping an infra-red cut-off and taking the on-shell
limit one obtains instead
$$\ln{\mu+\delta/2p_0\over\mu-\delta/2p_0} \to
{\delta\over p_0\mu}+O(\delta^3), $$
so that
$$\delta\Sigma(P)\Big|_{{\bf p}=0}
= {\xi g^2T\over2\pi^2\mu}{\delta\over p_0}\gamma_0 + O(\delta^2)
= {\xi g^2T\over2\pi^2\mu}\Sigma(P)\Big|_{{\bf p}=0} + O(\delta^2).
\eqno(7)$$
Evidently, there is no longer a contribution to the pole of the
full propagator, but only a (real!)
contribution to the {\it residue} $\propto T/\mu$.

At $T=0$, the residue of the fermion propagator is known to
be both gauge-dependent and infra-red singular, which enforces
the introduction of an infra-red cut-off to render
the propagator well-defined [\ref{rIZ}].
The only difference at finite temperature is
that the different infra-red behaviour due to
Einstein--Bose factors changes the singularity in the residue
from logarithmic to power-like.
At $T=0$, extraction of the gauge-independent mass counter-term
does not encounter the need of an infra-red regulator,
this becomes apparent only
when one goes on to compute also the residue.
But at finite temperature, without infra-red regularization,
the linear divergence in the residue is able to fake a
regular contribution
to the pole. With an infra-red cut-off, the correction to the
pole position at the resummed one-loop order, which gives the
leading order to the damping rate,
turns out to be both infra-red finite and gauge-independent.

In conclusion, the gauge-dependence problem of the resummation
scheme of Braaten and Pisarski, encountered by Baier et al.
in the leading-order evaluation of the damping rate of fermionic
excitations in high-temperature QED and QCD, is resolved by the
necessity of infra-red regularization of the on-shell expansion
of the resummed self-energy contributions.

\bigskip
I am indebted to Rolf Baier, Randy Kobes, Gabor Kunstatter,
Rob Pisarski, Dominique Schiff, and Hermann Schulz
for correspondence, as well as
Tanguy Altherr and Max Kreuzer for discussions.

\bigskip

\noindent{\bf References}
\newcounter{nom}
\begin{list}{[\arabic{nom}]}{\usecounter{nom}}
\item
For a review see e.~g.
R. Pisarski, Nucl. Phys. {\bf A525}, 175c (1991).
\label{rPr}
\item
R. Kobes, G. Kunstatter and A. Rebhan,
Phys. Rev. Lett. {\bf64}, 2992 (1990);
Nucl. Phys. {\bf B355}, 1 (1991).
\label{rKKR}
\item
The formal argument of Ref.~[\ref{rKKR}]
in favor of gauge independence
--- which does not refer to the
ill-defined [4] notion of an $S$-matrix at finite temperature ---
goes by relating the gauge variation
of a plasma dispersion relation,
say $a(p_0,|{\bf p}|)=0$, to itself times a certain Green function,
$\delta a(p_0,|{\bf p}|)=a(p_0,|{\bf p}|) \delta X(p_0,|{\bf p}|)$,
where the diagrammatic expansion of $\delta X$
is one-particle-irreducible,
except for Faddeev--Popov ghost propagation. At finite temperature,
ghost dispersion relations are uncorrelated to physical ones, hence
$\delta X$ is not expected to develop poles
simultaneously with $a=0$.
\label{fKKR}
\item
H. Narnhofer, M. Requardt and W. Thirring, Commun. Math. Phys.
{\bf 92}, 247 (1983).
\label{rNRT}
\item
R. D. Pisarski, Phys. Rev. Lett. {\bf63}, 1129 (1989);
E. Braaten and R. D. Pisarski,
Phys. Rev. Lett. {\bf64}, 1338 (1990);
Nucl. Phys. {\bf B337}, 569 (1990);
Phys. Rev. D {\bf42}, 2156 (1990).
\label{rBP}
\item
R. Baier, G. Kunstatter and D. Schiff,
Bielefeld preprint BI-TP 91/38, to appear in
Phys. Rev. Lett.
\label{rBKS}
\item
Such a partial resummation scheme
has been developed for the case of
gluonic excitations in:
M. Kreuzer, A. Rebhan and H. Schulz,
Phys. Lett. {\bf B244}, 58 (1990);
U. Kraemmer, M. Kreuzer, A. Rebhan
and H. Schulz, Lect. Notes in
Phys. {\bf 361}, 285 (1990).
\label{rKKRS}
\item
A finite $\lambda>0$, and with it gauge-parameter
independence, could be
provided by a modification of the
kernel $\delta_{ab}$ in the inhomogeneous gauge-breaking term
$(1/2\xi)(\partial_\mu A^\mu)^a\delta_{ab}(\partial_\nu A^\nu)^b$
according to
$\delta_{ab}\to\delta_{ab}
([\hbox{$\sqcup\llap{\hbox{$\sqcap$}}$}+\lambda]
/\hbox{$\sqcup\llap{\hbox{$\sqcap$}}$})^2,$
so that $\lambda$ enters as a second gauge parameter.
However, the Landau part of the gauge propagator ($\xi=0$) still
contains a massless double pole.
If the latter is again handled according to (4) and (5),
the result would still differ from
gauges such as the Coulomb or the
axial one which do not have a double pole.
\label{fG}
\item
In a recent paper, Nakkagawa et al.
[H. Nakkagawa, A. Ni\'egawa and
B. Pire, Palaiseau preprint CPTH-A092 (1992)]
have repeated the calculation
of Ref.~[\ref{rBKS}] with slight variations. Noting that the
gauge dependence of the result
is sensitive to changing limits like those in eq.~(5), they claim
the resolution of the gauge-dependence
problem. However, they choose
a procedure leading to gauge independence by fiat, without
any deeper justification.
\label{rNNP}
\item
C. Itzykson and J.-B. Zuber,
{\it Quantum Field Theory} (McGraw-Hill,
New York, 1980), ch.~7.1.2.
\label{rIZ}
\end{list}
\end{document}